# Enhanced superconductivity in atomically thin TaS$_2$

*Efrén Navarro-Moratalla\*[1†‡], Joshua O. Island\*[2‡], Samuel Mañas-Valero[1], Elena Pinilla-Cienfuegos[1], Andres Castellanos-Gomez[2], Jorge Quereda[3], Gabino Rubio-Bollinger[3,5], Luca Chirolli[4], Jose Angel Silva-Guillén[4], Nicolás Agraït[3,4,5], Gary A. Steele[2], Francisco Guinea[4], Herre S.J. van der Zant[2], and Eugenio Coronado\*[1]*

[1] Universidad de Valencia (ICMol), Catedrático José Beltrán Martínez nº 2, 46980, Paterna, Spain.

[2] Kavli Institute of Nanoscience, Delft University of Technology, Lorentzweg 1, 2628 CJ Delft, The Netherlands.

[3] Departamento de Física de la Materia Condensada. Universidad Autónoma de Madrid, Campus de Cantoblanco, 28049 Madrid, Spain.

[4] Instituto Madrileño de Estudios Avanzados en Nanociencia (IMDEA- Nanociencia), Calle Farady 9, Cantoblanco, 28049 Madrid, Spain.

[5] Condensed Matter Physics Center (IFIMAC), Universidad Autónoma de Madrid, 28049 Madrid, Spain.

**ABSTRACT**

The ability to exfoliate layered materials down to the single layer limit has opened the opportunity to understand how a gradual reduction in dimensionality affects the properties of bulk materials. Here we use this top-down approach to address the problem of superconductivity in the two-dimensional limit. The transport properties of electronic devices based on 2H tantalum disulfide flakes of different thicknesses are presented. We observe that superconductivity persists down to the thinnest layer investigated (3.5 nm), and interestingly, we find a pronounced enhancement in the critical temperature from 0.5 K to 2.2 K as the layers are thinned down. In addition, we propose a tight-binding model, which allows us to attribute this phenomenon to an enhancement of the effective electron-phonon coupling constant. This work provides evidence that reducing dimensionality can strengthen superconductivity as opposed to the weakening effect that has been reported in other 2D materials so far.

**INTRODUCTION**

The behavior of superconductors in the two-dimensional (2D) limit is a long-standing problem in physics that has been the focus of extensive research in the field.[1,2,3,4,5,6] The bottom-up approach has provided signs of the existence of superconductivity at the 2D limit in experiments performed on in situ grown, ultrathin lead films fabricated by evaporation.[7,8] However, for films grown in this way, it is difficult to avoid the strong influence from the substrate lattice, yielding typically highly disordered films. A different approach takes advantage of the ability of certain van der Waals materials to be separated into individual layers, which may later be isolated as defect-free 2D crystals on a substrate of choice.[9] This top-down approach permits overcoming the lattice and chemical restrictions imposed by the substrate in the bottom-up strategy in such a way that the coupling may be minimized by an appropriate choice of surface.[10,11,12]

Although graphene is not an intrinsic superconductor, recent studies have brought forward the possibility of inducing superconductivity in this 2D material by garnishing its surface with the right species of dopant atoms or, alternatively, by using ionic liquid gating.[13,14] However, reported experiments have failed to show direct evidence of superconducting behavior in exfoliated graphene, leaving out the archetypal material from studies of 2D superconductivity.[15]

An even more attractive family of 2D materials is provided by the transition metal dichalcogenides (TMDCs) since some of its members exhibit superconductivity in the bulk state.[16,17] Just as in graphene, TMDCs present a strong in-plane covalency and weak interlayer van der Waals interactions which allow exfoliation of the bulk.[18] This has given rise to a very rich chemistry of hybrid multifunctional materials based on the restacking of TMDC nano-layer sols with functional counterparts.[19,20] In addition, the all-dry exfoliation methodologies have allowed for the deposition of TMDC flakes on a variety of surfaces.[21,22] These micromechanical exfoliation techniques allow the access to nearly defect-free, large surface area flakes of virtually any TMDC, opening the door to the study of how a dimensionality reduction affects the properties of these materials.[23,24,25,26] Surprisingly, despite the works reported in the literature searching for intrinsic superconductivity in atomically-thin 2D crystals,[27,28,29] for a long time the sole examples came from FeSe thin films grown in situ on a substrate.[30,31,32,33] Only very recently, several studies of niobium diselenide ($NbSe_2$) flakes have yielded the first clear evidence of the existence of superconductivity in freshly cleaved specimens of less than three layers in thickness.[34,35,36,37]

Tantalum disulphide ($TaS_2$) is another member of the TMDC family. In its bulk state, $TaS_2$ is composed of robust covalently bonded S-Ta-S planes that stack upon each other. A variety of polytypic phases originate from the distinct in-plane Ta coordination spheres described by the $S^{2-}$ ligands and by the stacking periodicity of the individual planes. For instance, the 2H and 1T polytypes present unit cells with two trigonal bipyramidal and one octahedral Ta-coordinated layers, respectively. Though extensively explored in the 1960s,[38] 1T and 2H polytypes are once again attracting major attention since they constitute ideal case studies for the investigation of competing orders, namely, superconductivity, charge density waves (CDW),[39,40] and hidden phases.[41] In this scenario, the study of decoupled or isolated $TaS_2$ layers may provide new insights into these exotic phenomena.[42] Transport measurements of few-layer $TaS_2$ flakes have been reported in flakes as thin as 2 nm, but superconductivity in $TaS_2$ layers thinner than 8 nm has not been observed, probably due to the environmental degradation of the samples.[43]

Here, we explore 2D superconductivity in few-molecular-layer tantalum disulfide flakes of different thicknesses, which have been mechanically exfoliated onto $Si/SiO_2$ substrates. Interestingly, we observe that superconductivity persists down to the thinnest layer investigated (3.5 nm, approximately 5 covalent planes), with a pronounced increase in the critical temperature ($T_c$) from 0.5 K (bulk crystal) to approximately 2.2 K when the thickness of the layer is decreased. In search of the origin of these observations, we perform Density Functional Theory (DFT) calculations and construct a simple tight binding model to study the change in the electronic band structure and density of states (DOS) at the Fermi level as a function of reduced thickness. We ascribe the enhancement to an increase in the effective coupling constant ($\lambda_{\text{eff}}$) for reduced thicknesses, which ultimately determines $T_c$.

**RESULTS**

**Fabrication of transport devices**

While the exfoliation of other TMDC members has been extensively studied, little has been reported on the controlled isolation of atomically thin 2H-TaS$_2$ flakes. This layered material appears to be difficult to exfoliate and is also particularly susceptible to oxidation in atmospheric conditions,[44] hindering the manipulation of very thin flakes in open moist air. Though complex encapsulation techniques help preserving samples from oxidation,[34] we find that a rapid integration of freshly exfoliated flakes into final devices and their immediate transfer to vacuum conditions for measurement also permits retaining the pristine properties of most TaS$_2$ samples (vide infra).

The experimental process begins with the chemical vapor transport growth of bulk TaS$_2$ crystals (vide infra in Methods), which are subsequently exfoliated onto Si/SiO$_2$ substrates. To ensure a high-quality material, optical, Raman, and atomic force microscopy (AFM) characterization was performed on exfoliated flakes of varying thicknesses (see Supplementary Figures 1 – 5 and Supplementary Notes 1 and 2). As already established for graphene and other TMDCs, inspection of the substrate surface by optical microscopy permits identifying the presence of nanometer thin TaS$_2$ flakes. In an attempt to access flakes with a reduced number of atomic layers we developed a modification of the micromechanical exfoliation method and optimized it for the controlled isolation of few-layer 2H-TaS$_2$ flakes.[45,46] The method relies on precisely controlling a uniaxial pressure applied directly with a single crystal over the accepting substrate and in combination with a shearing cleavage movement. This allows for the cleavage of very thin flakes, down to 1.2 nm thick (see Supplementary Figure 1), corresponding to a single 2H-TaS$_2$ unit-cell (see Fig. 1a) formed by two individual layers. Unfortunately, all attempts to contact these flakes and measure transport properties were unsuccessful, likely due to their instability in ambient conditions.

To avoid the oxidation of few layer flakes, freshly exfoliated samples designated for device fabrication and transport measurements are immediately covered with an methyl

methacrylate/ poly-methyl methacrylate double layer resist in preparation for subsequent device nanofabrication steps. Figure 1b shows an example of a fabricated device incorporating the thinnest flake measured with a thickness of 3.5 nm (corresponding to approximately 5 layers) and lateral dimensions of the order of a few micrometers as imaged by AFM. The chromium/gold (Cr/Au, 5/70 nm) electrodes were evaporated onto selected flakes by employing standard e-beam lithography techniques (see Methods). All transport measurements were made using a four terminal current bias configuration in a temperature range of 20 mK to 4 K in a dilution fridge.

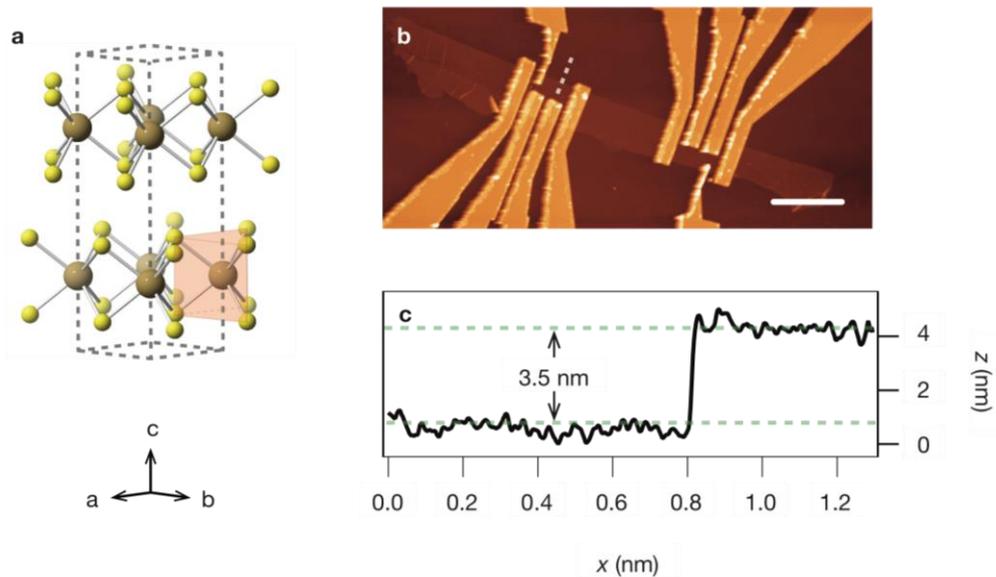

**Figure 1. Atomically-thin $TaS_2$ devices** (a) Ball and stick model of the crystal structure of the 2H polytype of $TaS_2$. The dashed prism encloses the content of a single unit cell and the metal coordination geometry is highlighted by the red polyhedron. (b) Atomic force microscopy (AFM) image of two devices fabricated on a 3.5 nm 2H-$TaS_2$ flake. The scale bar is 4 μm in length. The full color scale of the topograph corresponds to a height of 100 nm. (c) Line profile of the flake taken at the location of the white dotted line in (b).

**Transport properties and superconductivity** We present measurements on 12 flakes of varying thicknesses in the ≈ 3-30 nm range, integrated in the described four-terminal devices, with the aim of studying the effect of dimensionality reduction on the superconducting properties of $TaS_2$. All devices show a superconducting transition observed by four terminal current bias measurements as a function of temperature. Figure 2 shows the current-voltage (I-V) and resistance-temperature (R-T) characteristics for three representative devices having thicknesses of 14.9 nm (Figure 2(a) - (b)), 5.8 nm (Figure 2(c) - (d)), and 4.2 nm (Figure 2(e) - (f)). The transport data for the thinnest 3.5 nm flake can be found in the Supplementary Figure 6. The zero bias, numerical derivatives (dV/dI) as a function of temperature show a clear superconducting transition for each device (Figures 2(b), (d), and (f)). From these (interpolated) curves we estimate $T_c$, taken at 50% of the normal-state resistance. For the 14.9 nm flake, and despite the fact that the sample does not attain a zero resistive state, one may still appreciate that there is a phase transition centered at 540 ± 230 mK. This is in rough agreement with previously reported $T_c$ values of 600 mK for bulk 2H-$TaS_2$ material.[47] Interestingly, and in contrast with studies on other 2D superconductors, the $T_c$ values show a marked increase for the thinner flakes of 5.8 nm (1.45 ± 0.13 K) and 4.2 nm (1.79 ± 0.20 K). This peculiar result is discussed in detail below. Additionally, critical current densities increase by orders of magnitude as the devices become thinner (14.9 nm, $J_c$ ≈ 700 A $cm^{-2}$, 5.8 nm, $J_c$ ≈ $7 \times 10^4$ A $cm^{-2}$, and 4.2 nm, $J_c$ ≈ $5 \times 10^5$ A $cm^{-2}$). In thin film superconductors with high critical current densities, as those measured in our thinnest flakes, Joule self-heating starts to play a role.[48] This explains the pronounced asymmetry in the I-V characteristics for thinner flakes (Figure 2(a) versus Figure 2(e)). As the current bias is swept from high negative values through zero, non-equilibrium Joule heating pushes the superconducting transition to a lower current value. This asymmetry decreases as the temperature approaches $T_c$, where Joule heating effects become less significant (Figure 2(e)).

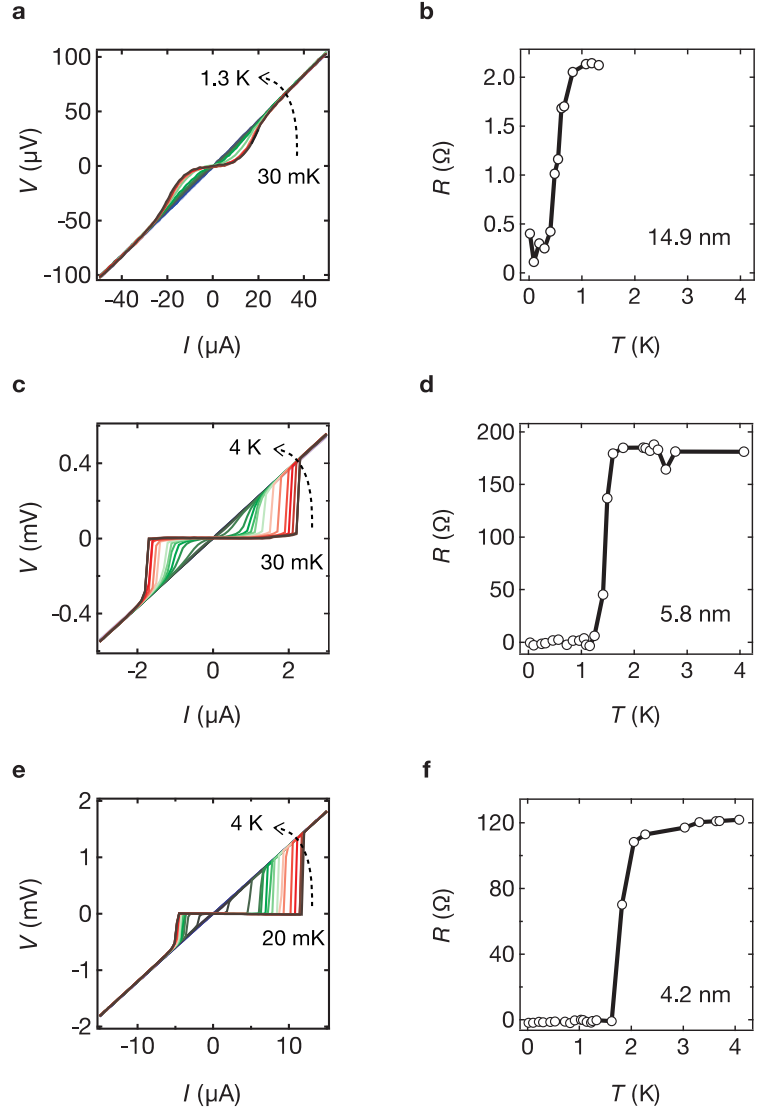

**Figure 2. Superconductivity in atomically thin crystals.** Temperature dependence of three selected devices spanning the range of thicknesses studied. (a) Current-voltage (I-V) characteristics as a function of temperature for a bulk-like 14.9 nm device. (b) Resistance (zero bias numerical derivative) vs. temperature for the 14.9 nm device. (c) I-V characteristics as a function of temperature for a 5.8 nm device. (d) Resistance vs. temperature for the 5.8 nm device. (e) I-V characteristics as a function of temperature for a 4.2 nm device. (f) Resistance vs. temperature for the 4.2 nm device.

**Effect of an external magnetic field on superconductivity**

To further characterize the devices at 50 mK, the upper critical field ($B_{c2}$) of these type II superconductors is determined by applying an external magnetic field, perpendicular to the surface of the flake. Figure 3 shows color scale plots of d$I$/d$V$-$I$ curves as a function of external field for the same three devices as in Figure 2. Figure 3(b) shows the zero-bias differential resistance as a function of external field. From these curves we estimate the $B_{c2}$ as the external field at which the device returns to the normal-state resistance. Once again, in accordance with the upper critical field reported for the bulk material (110 mT), we measure a $B_{c2}$ of ≈ 130 mT for the bulk-like 14.9 nm flake.[47] The thinner flakes present higher upper critical fields of ≈ 0.9 T (5.8 nm) and ≈ 1.7 T (4.2 nm) following the interesting trend for $T_c$. The critical fields at 50 mK allow estimation of the superconducting Ginzburg-Landau coherence lengths given by: $B_{c2}(50 \text{ mK}) = \varphi_0/2\pi\xi(50 \text{ mK})^2$. The coherence lengths for the 4.2 nm and 5.8 nm flakes are 13.9 nm and 19.1 nm, respectively, suggesting that these flakes are in the 2D limit. To further qualify the 2D nature of the thinnest flakes, we analyze the $I$-$V$ and $R$-$T$ curves (such as those in Figure 2) of selected devices at zero external field in order to infer the typical signature of 2D superconductivity: the Berezinskii–Kosterlitz–Thouless (BKT) transition (see Supplementary Figure 7 and Supplementary Note 3). Note that this study can only be carried out for selected thinner samples for which sufficient data is available. We find that the transport data is consistent with a BKT superconducting transition, further supporting the 2D nature of the thinnest TaS$_2$ flakes.

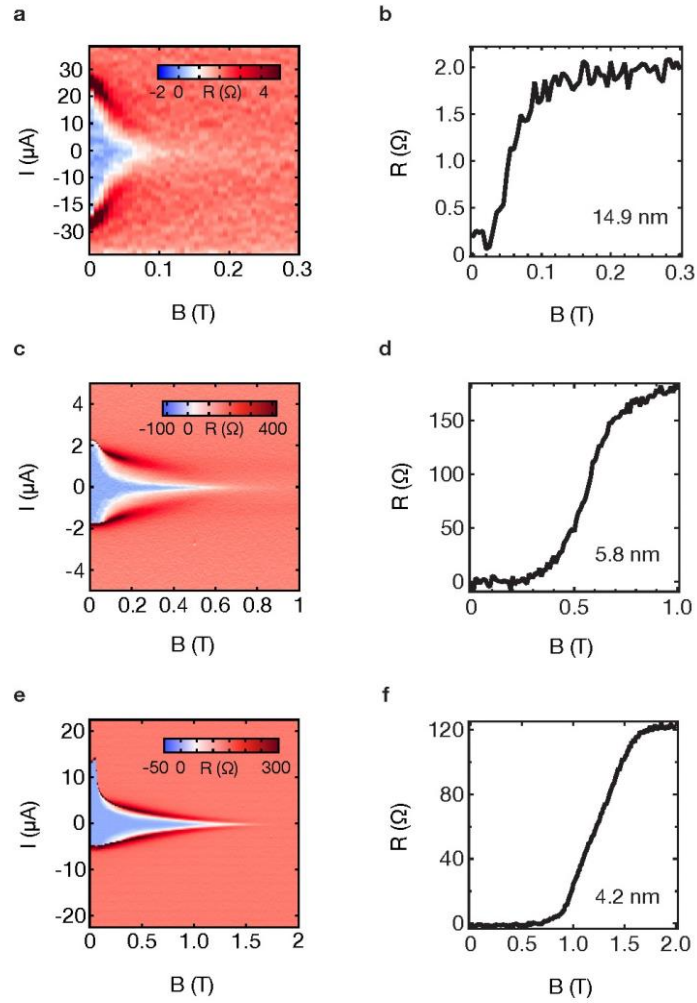

**Figure 3. Enhanced critical magnetic field in thin flakes.** Perpendicular external magnetic field dependence at 30 mK for three selected devices spanning the range of thicknesses studied. (a) Resistance (zero bias numerical derivative) vs. applied field for a bulk-like, 14.9 nm device. (b) Zero bias resistance vs. applied field for the 14.9 nm device. (c) Resistance (zero bias numerical derivative) vs. applied field for the 5.8 nm device. (d) Zero bias resistance vs. applied field for the 5.8 nm device. (e) Resistance vs. applied field for the 4.2 nm device. (f) Zero bias resistance vs. applied field for the 4.2 nm device.

**Effect of dimensionality on the superconducting state**

We now turn our attention to the collective behavior of our 12 devices and the effect of reduced dimensionality on the superconducting properties of TaS$_2$. Figure 4 illustrates the measured $T_c$ and $B_{c2}$ for the devices reported. A bulk limit was found for samples over 10 nm in thickness, such as the one in figures 2(a - b) and 3(a - b), for which the superconducting properties were consistent with bulk crystals and did not depend on the number of layers. It is interesting to note that these types of flakes exhibit a non-zero residual resistance (red data points) at base temperature, indicating a certain degree of crystalline inhomogeneity and providing a plausible explanation to the slight variation of $T_c$, similar to the variation in reported bulk values (0.6 K and 0.8 K).[25,49] The bulk-limit devices approach the edge of the 2D limit set by the Ginzburg-Landau (GL) coherence length ($\xi = 55$ nm) estimated from the bulk $B_{c2}$ (see Figure 4b).

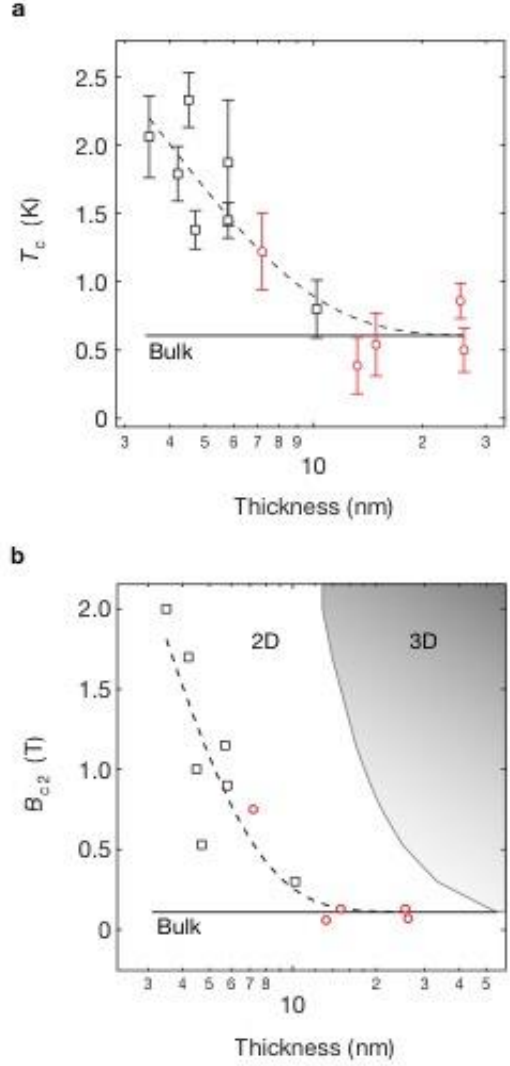

**Figure 4. 2D superconductivity and enhanced $T_c$ in atomically thin TaS$_2$** (a) Variation of $T_c$ as a function of the thickness of the TaS$_2$ flakes. Devices exhibiting a non-zero residual resistance below $T_c$ are plotted in red. The error bars are given by the temperatures at 10% and 90% of the normal state resistance. The solid black line marks the bulk $T_c$ of 600 mK. The black dotted line is an exponential trend line, fit to the data starting at the bulk limit. (b) Variation of $B_{c2}$ as a function of flake thickness. The red circles mark the same devices in (a) having residual resistance. The black solid line indicates the bulk limit upper critical field of 110 mT. The grey solid line plots the GL coherence lengths, calculated from the y-axis $B_{c2}$ values, and marks the edge of the 2D limit.

**DISCUSSION**

In addition to thicker flakes that behave in a way consistent with bulk properties, we also observe the superconducting transition in devices made out of thinner TaS$_2$ flakes, down to 3.5 nm (~ 5 layers). Interestingly, we observe a strong enhancement of $T_c$ and $B_{c2}$ for thinner flakes, up to more than a factor of four larger than in the bulk material. The $T_c$ enhancement with decreasing number of layers exhibited by the TaS$_2$ samples is in strict contrast to the $T_c$ suppression previously reported in elemental materials,[7] binary systems,[50] and even the closely related dichalcogenide family member, NbSe$_2$.[27] A common theme in these studies is that as the material is thinned down, substrate interactions, either from induced strain or increased Coulomb interactions, suppress the formation of Cooper pairs. In NbSe$_2$ devices, a clear correspondence can be made with a decrease in the residual resistance ratio (RRR) giving an indication of increased substrate interactions or more probable that flake degradation is more prevalent in thinner flakes.[28] This agrees with our attempts to contact flakes thinner than 3.5 nm showing a complete insulating state at room temperature. Correspondingly, the RRR values (see Supplementary Figure 8) for our TaS$_2$ sample set show a significant reduction for the two thinnest flakes. However, devices as thin as 4.5 nm still maintain an RRR of 10, indicating pristine thin samples even below our bulk limit of 10 nm.

An initial point that needs to be addressed once trying to interpret the $T_c$ enhancement is the possibility of electrochemical doping coming from either the original crystals, or through fabrication processes (lithography resists). Although it is well understood that the $T_c$ of TaS$_2$ crystals is particularly sensitive to intrinsic non-stoichiometric doping,[51] we may rule out this effect coming from the original crystals by having measured a bulk $T_c$ of approximately 0.5 K. Now considering the potential doping coming from environmental or intercalation interactions, Raman spectroscopy provides us with strong evidence of the absence of such processes. In contrast with the remarkable peak shifts displayed by intercalated crystals of 2H-MX$_2$,[52] the Raman spectra of exfoliated flakes presented in the supplement (review Supplementary Figures 3 – 5) show no significant change in crystal structure for flakes of only 4 layers. Given that the flakes are not undergoing intercalation through exfoliation or fabrication, there could indeed be some doping coming from surface contamination or from the oxide substrate. However, previous studies show that gate-induced or surface-induced electrostatic doping allows for a carrier density modulation of maximum ca. $10^{12}$ cm$^{-2}$,[53] which is at least three orders of magnitude lower than the estimated single layer carrier concentration in these metallic TMDCs (ca. $10^{15}$ cm$^{-2}$).[28,54] In this line, these doping effects have shown to modulate $T_c$ in NbSe$_2$ by 8% at most.[28] Finally, while substrate interactions have led to the interesting $T_c$ enhancements found in epitaxial grown FeSe on STO, we rule out such effects as the TaS$_2$ flakes presented here are weakly coupled to the substrate. This suggests a deeper mechanism as opposed to simple substrate interaction, intercalation, or degradation reported in previous studies.

A possible mechanism at work could be the enhancement of the superconducting properties associated with a suppression of the commensurate charge density wave (CCDW) order, which is in direct competition with superconducting pairing.[17] This is consistent with the interpretation presented of the enhanced $T_c$ and $B_{c2}$ observed in the studies of intercalation of TMDC, where it is argued that the in-plane chemical doping leads to the suppression of the charge density order, and in certain TMDCs under pressure where the same claim is made.[55,56,57] In order to explore the effect of the CDW on the DOS at the Fermi level as a function of reduced thickness, we calculate the DOS from an effective one-orbital tight-binding model and simulate the CDW at a mean field level as a periodic potential that locally shifts the onsite energy (see Supplementary Figures 9 and 10 and Supplementary Note 4). We find that the DOS at the Fermi level is not appreciably affected by the CDW for reduced thicknesses. Ultimately, to determine if such a competition with CDWs could be playing a role, one could search for direct evidence of such suppression in STM studies of thin flakes below the 10 nm bulk limit observed here.

An alternative explanation of the enhanced $T_c$ could be a change of the band structure of the material in atomically thin flakes. To explore this possibility, we perform DFT calculations and construct a simplified tight-binding model to study the electronic band structure and density of states (DOS) $\nu_N(0)$ as a function of the sample thickness. The results of the calculation can be observed in the Supplementary Figures 11 and 12 and Supplementary Notes 5 and 6. The resulting two-dimensional bands contain hole pockets and show saddle points below the Fermi level. These saddle points give rise to van Hove peaks, whose height increases as the number of layers is decreased, and ultimately diverge in the 2D limit. However, the density of states per layer at the Fermi level $\nu_N(0)$ decreases as the number of layers is reduced (see Supplementary Figure 13). For a simplified model with a constant attractive interaction $V$, the coupling constant, that ultimately determines the $T_c$, takes the usual BCS value $\lambda = V\nu_N(0)$. This behavior of the DOS would suggest at first an analogous trend of $T_c$, which does not suffice to explain the experiments. The value of the superconducting gap and $T_c$ can be influenced by the interactions properties of the material. The effective coupling constant[58] determining $T_c$ is given by $\lambda_{\text{eff}} = \lambda - \mu^*$, where $\lambda$ is the electron-phonon coupling constant, and $\mu^*$, known as Anderson-Morel pseudo-potential, is a term that represents the renormalized repulsive Coulomb interaction. In usual 3D superconductors characterized by a featureless -hence constant- DOS, the projection on the Fermi level of the high energy degrees of freedom gives rise to a pseudo-potential of the form $\mu^* = \mu/(1 + \mu \ln(W/\omega_0))$, with $\omega_0$ the characteristic phonon frequency, $W$ the system bandwidth, and $\mu$ the bare Coulomb repulsion. In a 2D system, with a DOS characterized by a van Hove singularity near the Fermi level, the renormalization of the bare $\mu$ can be significantly larger than in a 3D material. This effect is strongly dependent on the number of layers. For a generic DOS $\nu_N(\varepsilon)$, the pseudo-potential takes the form

$$\mu^* = \frac{\mu}{1 + \mu \int_{\omega_0}^{W} d\varepsilon \, \tilde{\nu}_N(\varepsilon)/\varepsilon}$$

with $\tilde{\nu}_N(\varepsilon)$ the total DOS normalized with its value at the Fermi energy. Assuming a constant repulsive interaction $U$, one can estimate $\mu = U\nu_N(0)$. For a weak repulsion, the renormalization is negligible and the effective coupling constant follows the DOS at the Fermi level $\nu_N(0)$. For a relatively strong Coulomb repulsion, the value of the pseudo-potential at the Fermi level can be strongly affected by features of the DOS at higher energies, such as van Hove singularities. As the number of layers is decreased, the renormalization of a relatively strong repulsion for the band structure in the model is sufficient to reverse the dependence of $T_c$ on the number of layers obtained from a simple electron-phonon attractive interaction (see Supplementary Figure 14). This analysis points to a non-negligible role of the Coulomb repulsive interaction in superconducting 2H-TaS$_2$, characterized by a predominant Ta 5d orbital character at the Fermi level. The Coulomb repulsion has also been proposed to be at the origin of superconductivity in MoS$_2$.[59,60,61]

In conclusion, we have reported 2D superconductivity in 2H-TaS$_2$ in atomically thin layers. In contrast to other van der Waals superconductors such as NbSe$_2$, we find that the $T_c$ of this material is strongly enhanced from the bulk value as the thickness is decreased. In addition to a possible charge-density wave origin, we propose a model in which this enhancement arises from an enhancement of the effective coupling constant, which determines the $T_c$. Our results provide evidence of an unusual effect of the reduction of dimensionality on the properties of a superconducting 2D crystal and unveil another aspect of the exotic manifestation of superconductivity in atomically thin transition metal dichalcogenides.

## METHODS

### Crystal growth

Polycrystalline 2H-TaS$_2$ was synthesized by heating stoichiometric quantities of Ta and S in an evacuated quartz ampoule at 900 ºC for 9 days. The growth of large single crystals from the polycrystalline sample was achieved by employing a three-zone furnace. The powder sample was placed in the leftmost zone of the furnace and the other two zones were initially brought to 875 ºC and kept at that temperature for 1 day. Following, the temperature of the source zone was risen to 800 ºC during the course of 3 hours. The temperature of the centre zone was then gradually cooled down at a speed of 1 ºC min$^{-1}$ until a gradient of 125 ºC was finally established between the leftmost (875 ºC) and centre (750 ºC) zones. A gradient of 50 ºC was also set between the rightmost and growth zones. This temperature gradient was maintained for 120 hours and the furnace was then switched off and left to cool down naturally. The crystals were then thoroughly rinsed with diethyl ether and stored under an N$_2$ atmosphere.

### Device fabrication

Contact pads and optical markers are first created on the surface of the Si/SiO$_2$ substrates to locate and design contacts to the transferred flakes. The contacts (chromium−5 nm/ gold−70 nm) are then patterned with standard e-beam lithography (Vistec, EBPG5000PLUS HR 100), metal deposition (AJA International), and subsequent lift-off in warm acetone. In order to preserve the sample integrity, it is crucial to exfoliate, pattern the electrodes, and load into the dilution fridge within a few hours. In that respect and even after minimizing the fabrication time, all attempts to contact flakes with thicknesses below 3.5 nm were unsuccessful due to sample degradation.

### Band structure calculations

The DFT simulation of the band structure of 2H-TaS$_2$ has been performed using the Siesta code on systems with different number of layers[62] We use the generalized gradient approximation (GGA), in particular, the functional of Perdew, Burke and Ernzerhoff.[63] In addition, we use a split-valence double-$\zeta$ basis set including polarization functions.[64] The energy cut-off of the real space integration mesh was set to 300 Ry and the Brillouin zone $k$ sampling was set, within the Monkhorst-Pack scheme,[65] to 30 x 30 x 1 in the case of multi-layer samples and 30 x 30 x 30 in the case of the bulk calculation. We use the experimental crystal structure of 2H-TaS$_2$ for all the calculations.[66]

**SUPPORTING INFORMATION**

Supporting Information contains a more detail explanation about: optical characterization, Raman characterization, transport properties and the residual resistance ratio (RRR) of different thickness devices, BKT fits for selected devices, DFT calculations and tight-binding model and mechanism leading to an enhanced effective coupling constant.

**AUTHOR INFORMATION & CONTRIBUTIONS**

*\* Corresponding Authors*

Dr. Efren Navarro-Moratalla

Universidad de Valencia (ICMol)

Catedrático José Beltrán Martínez nº 2, 46980, Paterna (Spain)

E-mail: enavarro@mit.edu

Joshua Island

E-mail: j.o.island@tudelft.nl

Prof. Eugenio Coronado

E-mail: eugenio.coronado@uv.es

*Present Addresses*

† Dr Efren Navarro-Moratalla

Department of Physics, Massachusetts Institute of Technology, 77 Massachusetts Avenue, Cambridge, Massachusetts 02139, United States.

Dr. Elena Pinilla Cienfuegos
Valencia Nanophotonics Technology Center, Building 8F | 1st Floor, Universidad Politécnica de Valencia, Camí de Vera, s/n, 46022 Valencia, Spain.

Dr Andres Castellanos-Gomez

Instituto Madrileño de Estudios Avanzados en Nanociencia (IMDEA- Nanociencia), Calle Farady 9, Cantoblanco 28049 Madrid, Spain.

*Author Contributions*

‡ These authors contributed equally to this work.

The manuscript was written through contributions of all authors. All authors have given approval to the final version of the manuscript. E.N.-M. jointly conceived the study with J.O.I., designed and performed the measurements, and prepared the manuscript; S.M.-V. and E.P.-C. helped with the preparation and characterization of samples and contributed to the edition of the manuscript. L.C. created the theoretical model with contributions from J.A.S.-G. and F.G. supervised the theoretical analysis and edited the manuscript. J.Q. and G. R.-B. optical microcopy measurements model and interpret. A.C.-G., N. A., G.A.S., H.S.J. vd Z., and E.C. supervised the study and edited the manuscript.

**COMPETING FINANCIAL INTERESTS STATEMENT**

The authors declare no competing financial interest

**ACKNOWLEDGEMENTS**

Financial support from the EU (ELFOS project and ERC Advanced Grant SPINMOL), the Spanish MINECO (Excellence Unit "Maria de Maeztu" MDM-2015-0538, Project Consolider-Ingenio in Molecular Nanoscience and projects MAT2011-25046 and MAT2014-57915-R, co-financed by FEDER), Dutch organization for Fundamental Research on Matter (FOM), NWO/OCW, and the Comunidad Autonoma de Madrid (MAD2D-CM -S2013/MIT-3007- and NANOFRONTMAG-CM -S2013/MIT-2850) and the Generalitat Valenciana (Prometeo

Program) are gratefully acknowledged. AC-G acknowledges financial support from the BBVA Foundation through the fellowship "I Convocatoria de Ayudas Fundacion BBVA a Investigadores, Innovadores y Creadores Culturales" ("Semiconductores ultradelgados: hacia la optoelectronica flexible"), from the MINECO (Ramón y Cajal 2014 program, RYC-2014-01406) and from the MICINN (MAT2014-58399-JIN). The authors are grateful to the Electronic Microscopy team at Central Support Service in Experimental Research (SCSIE, University of Valencia, Spain) for their kind and constant support.**REFERENCES**

[1] Beasley, M. R., Mooij, J. E., & Orlando, T. P. Possibility of Vortex-Antivortex Pair Dissociation in Two-Dimensional Superconductors. *Phys. Rev. Lett.* **42**, 1165–1168 (1979)

[2] Haviland, D. B., Liu, Y. & Goldman, A. M Onset of superconductivity in the two-dimensional limit. *Phys. Rev. Lett.* **62**, 2180–2183 (1989)

[3] Yazdani, A. & Kapitulnik, A. Superconducting-Insulating Transition in Two-Dimensional α-MoGe Thin Films. *Phys. Rev. Lett.* **74**, 3037–3040 (1995)

[4] Goldman, A. M. & Markovic, N. Superconductor-insulator transitions in the two-dimensional limit. *Phys. Today* **11**, 39–44 (1998)

[5] Guo, Y., et al. Superconductivity Modulated by Quantum Size Effects. *Science* **306**, 1915–1917 (2004)

[6] Hermele, M., Refael, G., Fisher, M. P. A. & Goldbart, P. M. Fate of the Josephson effect in thin-film superconductors. *Nature Phys.* **1**, 117–121 (2005)20

**SUPPLEMENTARY FIGURES**

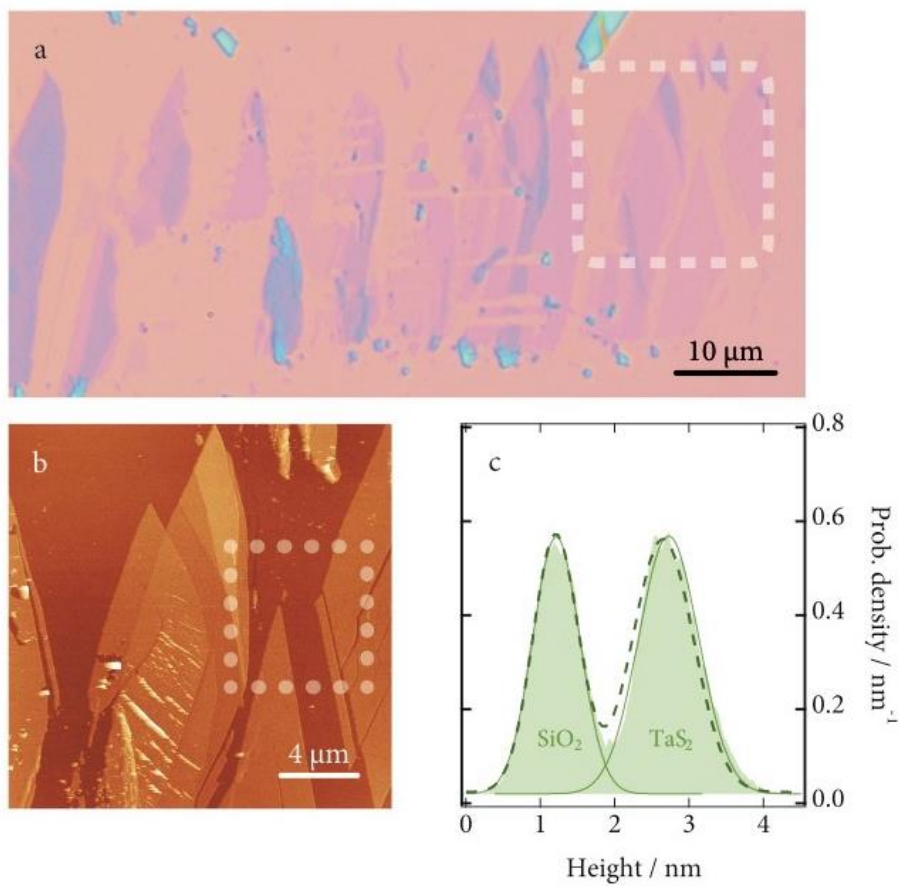

**Supplementary Figure 1. Atomically thin TaS$_2$ flakes deposited on a Si/285 nm SiO2 substrate by the optimised press and shear micromechanical exfoliation method.** [a] Optical microscopy image of a region of the substrate displaying a high density of atomically thin flakes. [b] AFM image of the region highlighted in a by the dashed box. [c] Probability density distribution of heights inside the dotted box in b. In this particular image a flake thickness of 1.2 ± 0.5 nm may be estimated.

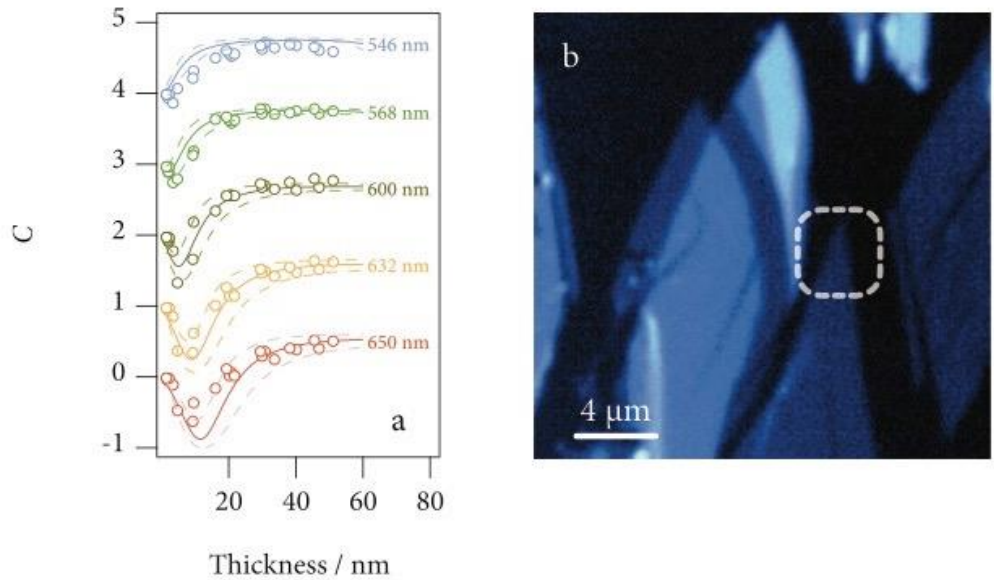

**Supplementary Figure 2. Optical contrast study of 2H-TaS$_2$ flakes.** [a] Optical contrast ($C$) of a selection of flakes as a function of their thickness under five different monochromatic illumination wavelengths ($\lambda$) between 546 nm and 650 nm. The solid lines correspond to the Fresnel-law-model calculation using the refractive index reported in the literature.[7] Note that the sets of contrast measurements for different $\lambda$ have been shifted vertically by 1, 2, 3 and 4 units for clarity. The uncertainty in $C$ due to a ± 10% variation in the real and imaginary parts of the refractive index is indicated by the dashed lines. [b] Optical contrast image at $\lambda$ = 600 nm TaS$_2$ flakes. The flake region marked by the dashed box is 1.2 nm thick, measured by AFM in contact mode, and shows a negative optical contrast of -0.03.

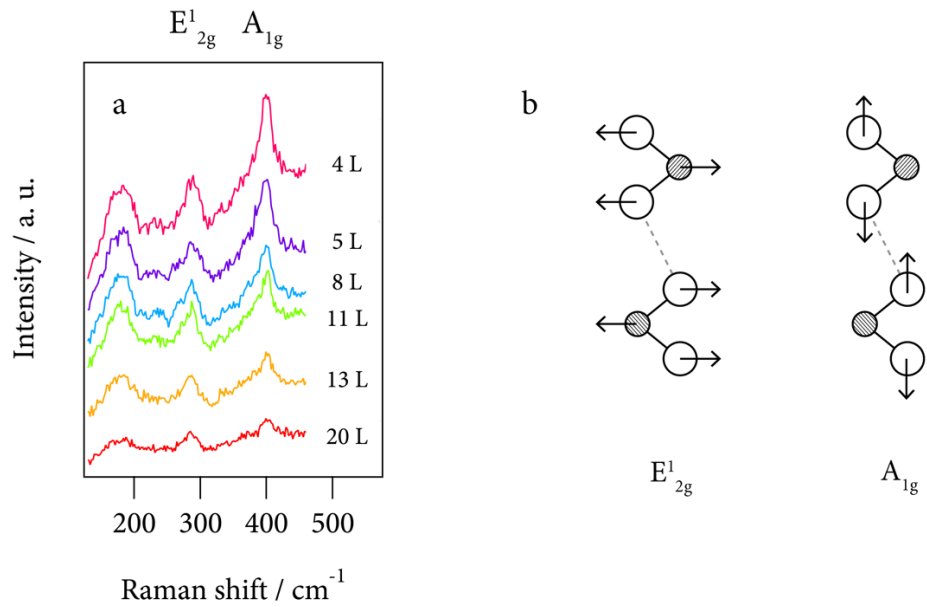

**Supplementary Figure 3. Raman spectroscopy of 2H-TaS$_2$ flakes.** [a] Raman spectra measured for 2H-TaS$_2$ flakes with thickness ranging from four layers to 20 layers. [b] Schematic representation of the vibration modes that correspond to the most prominent peaks at [a].

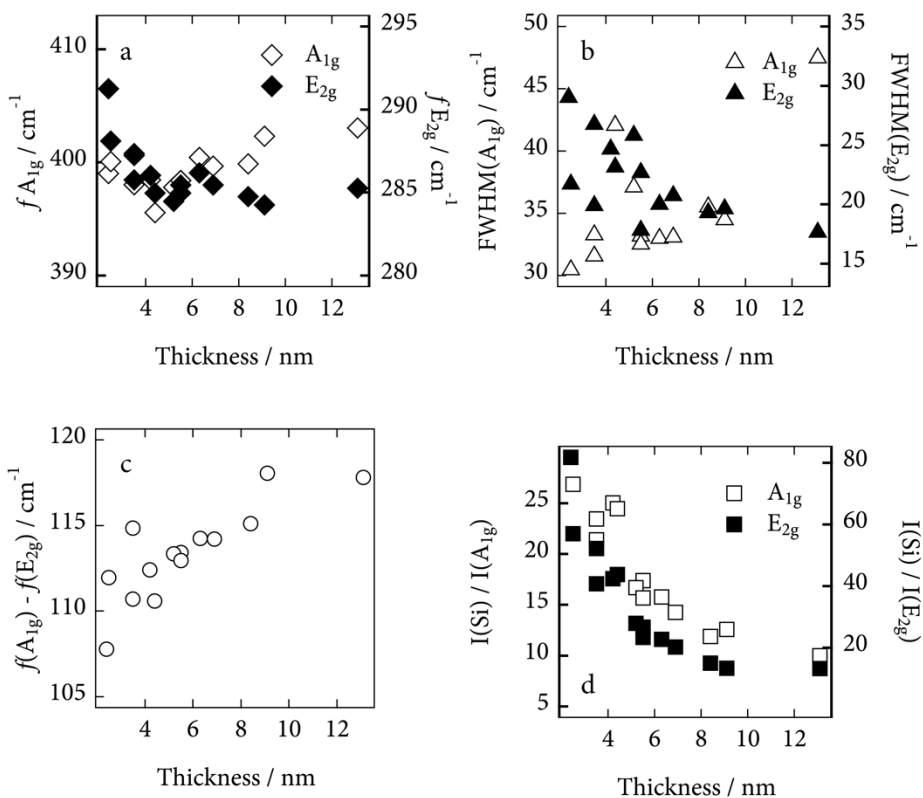

**Supplementary Figure 4. Thickness dependence of different Raman features of a selection of 2H-TaS$_2$ flakes.** [a] Frequency shift; [b] FWHM of the $A_{1g}$ and $E_{2g}$ Raman modes; [c] frequency difference between the $A_{1g}$ and $E_{2g}$ Raman modes; and [d] Raman intensity ratio between the Si peak (at 521 cm$^{-1}$) and the $A_{1g}$ and $E_{2g}$ peaks.

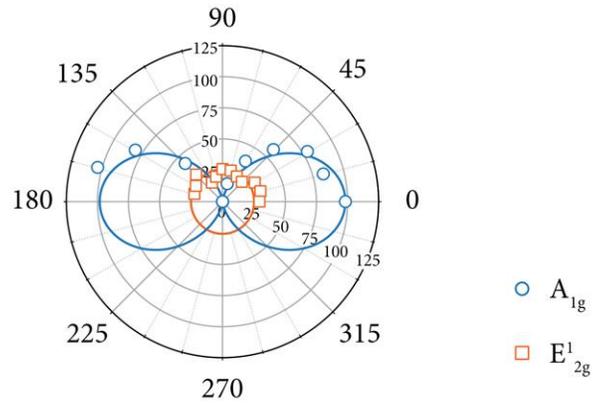

**Supplementary Figure 5. Angular dependence of the 2H-TaS$_2$ signal.** Intensity of the A$_{1g}$ and E$_{2g}$ Raman peaks as a function of the angle (in degrees) between linearly polarized excitation and detection.

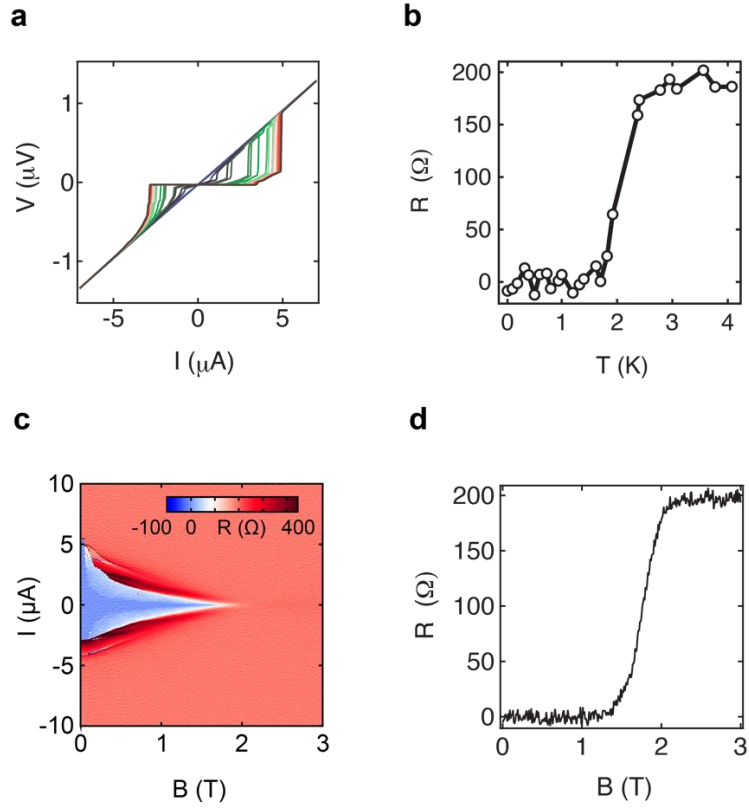

**Supplementary Figure 6. Transport properties of a device made out of a 3.5 nm thick TaS$_2$ flake.** [a] Current-voltage (I-V) characteristics as a function of temperature [b] Resistance (zero bias numerical derivative) vs. temperature curve. [c] Resistance vs. applied field and bias current. [d] Zero bias resistance vs. applied field.

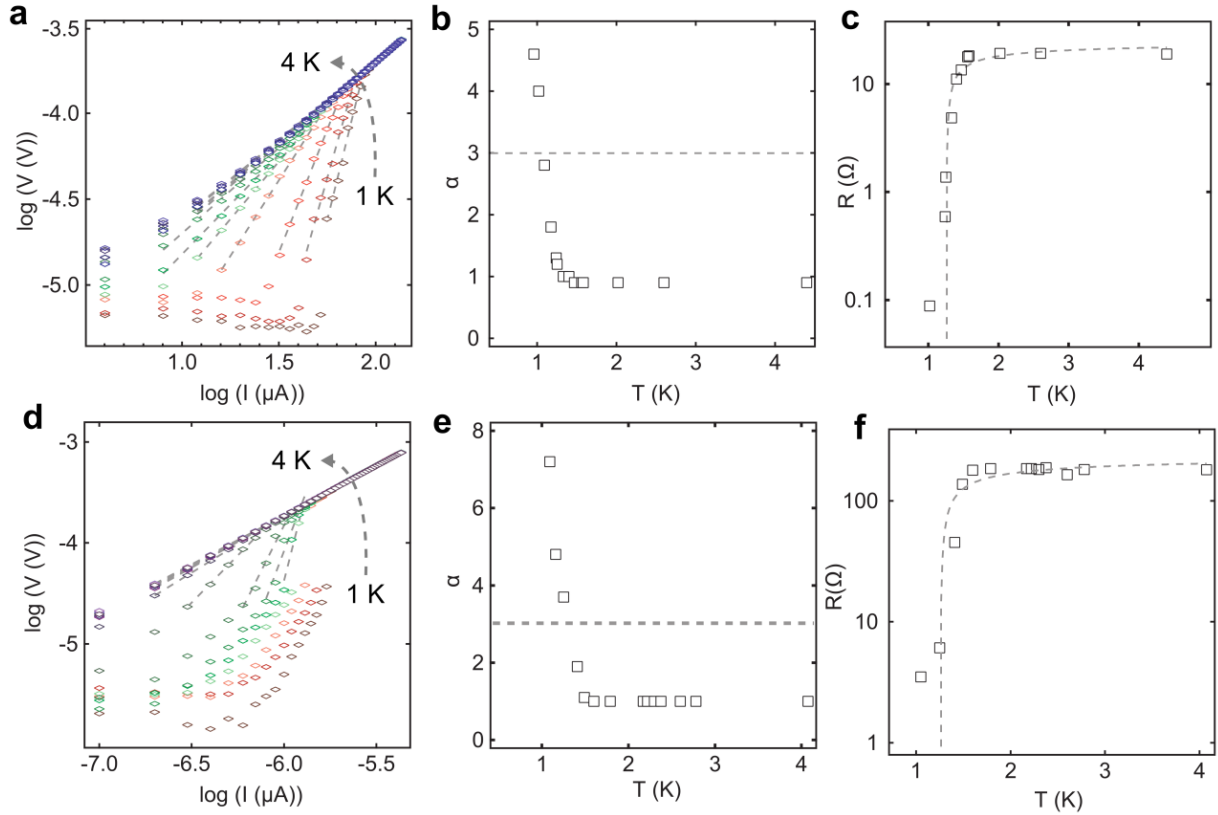

**Supplementary Figure 7. Example of a BKT fit performed for a 4.7 nm [a - c] and 5.8 nm [d - e] thick sample.** [a, d] *I-V* curves are displayed in a log-log scale. [b, e] The variation of the α parameter with temperature as a function of temperature, where the α = 3 value is indicated by the black dashed line and corresponds to $T_{BKT}$. [c, f] *R-T* curves as a function of temperature. The black dotted line plots the $R(T)$ dependence of the BKT model (see main text).

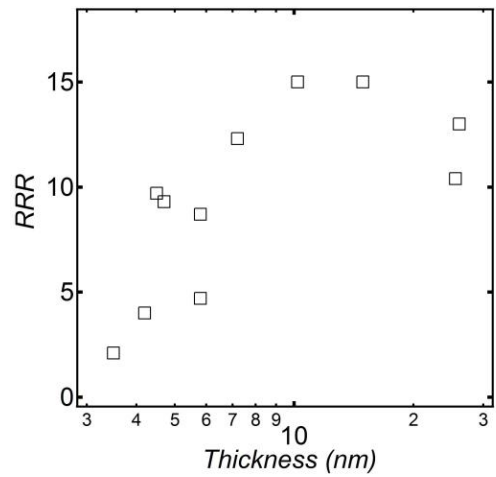

**Supplementary Figure 8. Resistance Residual Ratio (*RRR*) as a function of flake thickness.** The *RRR* is calculated as the ratio between the room temperature (297 K) resistance and the low temperature resistance at 4 K (*RRR* = *R*(297K)/*R*(4K)). High *RRR* values (~10) are still maintained below the bulk limit thickness of 10 nm indicating pristine flakes and absence of strong substrate interaction.

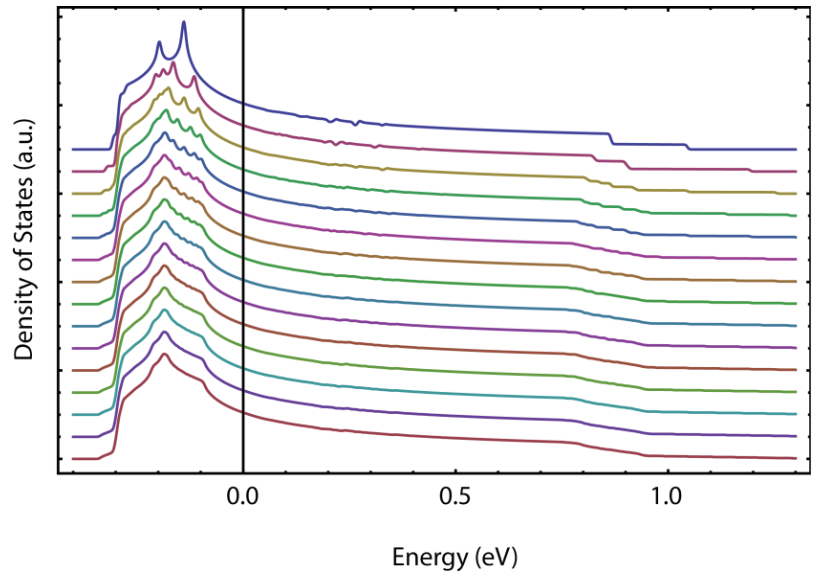

**Supplementary Figure 9. Density of states for N=1,…,15 layer 2H-TaS$_2$ systems in presence of a CDW potential.**

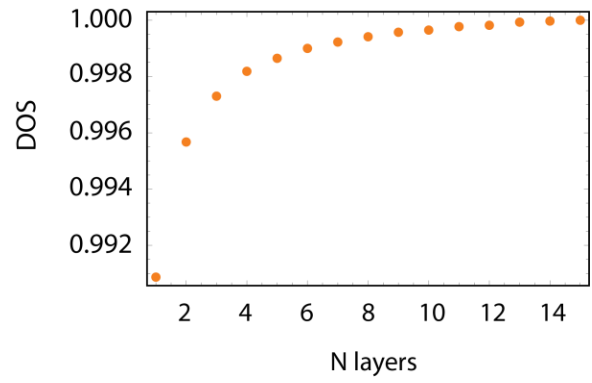

**Supplementary Figure 10. Density of states (DOS) at the Fermi level versus N layers of 2H-TaS$_2$ in presence of the CDW modulation.**

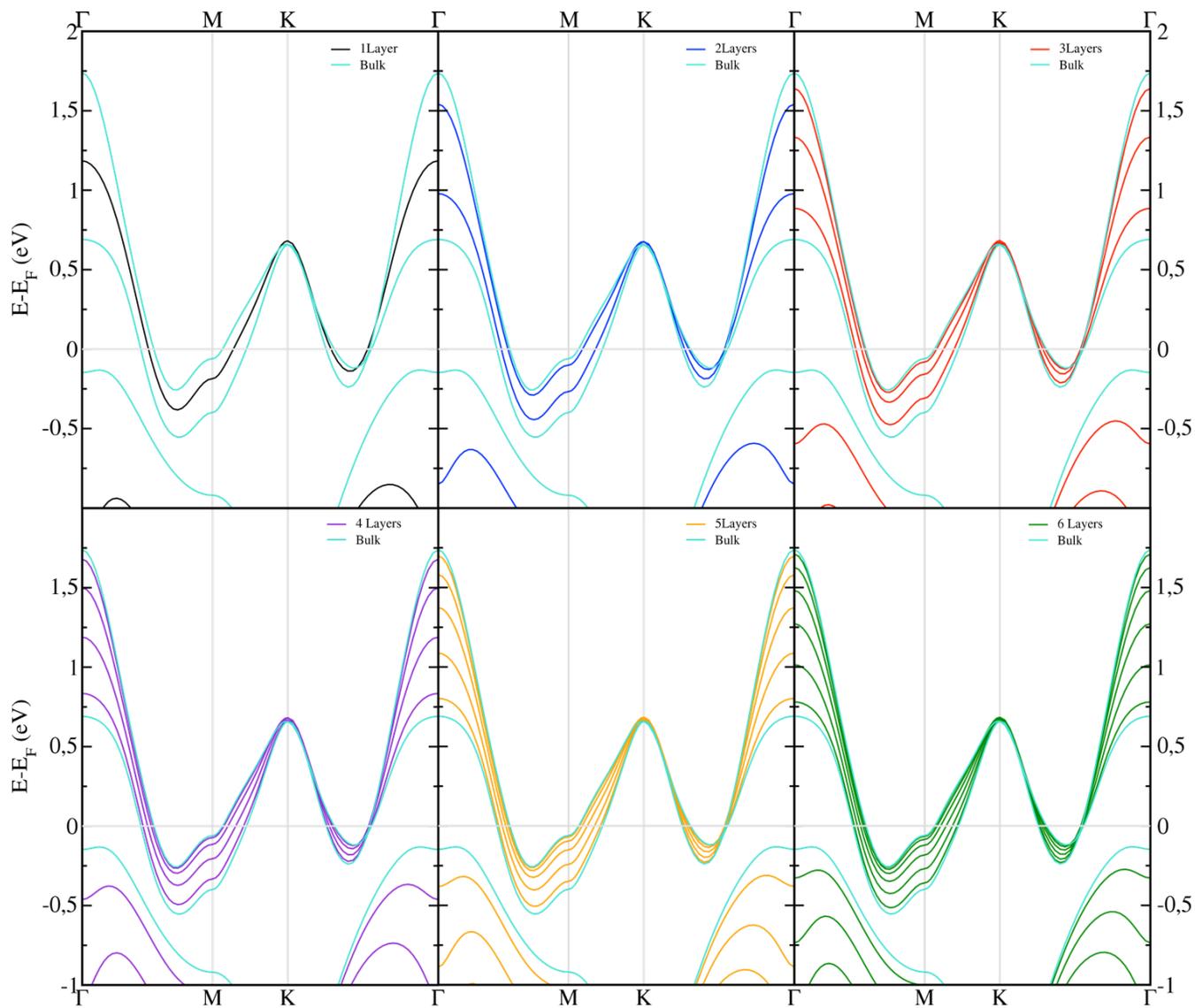

**Supplementary Figure 11. DFT band structure of the different systems with varying number of 2H-TaS2 layers from 1 to 6.** The bulk band structure (light blue) is plotted in all charts as a reference.

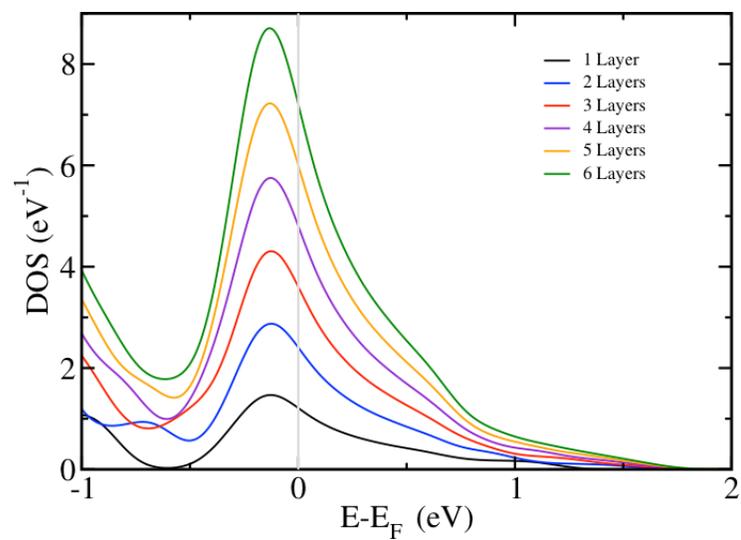

**Supplementary Figure 12. Non-normalized DFT calculated density of states (tot-DOS) with varying number of 2H-TaS2 layers.**

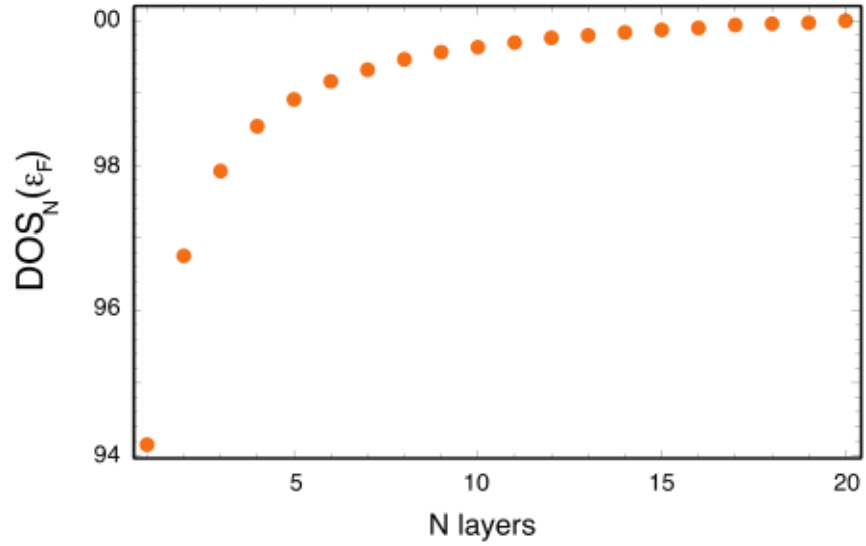

**Supplementary Figure 13. Relative change of the Density of States per layer at the Fermi level $v_N(0)$ with varying number of TaS$_2$ layers.** The DOS is obtained from the tight-binding model, and the DOS of the *N*-th layer is normalized to the one with 20 layers.

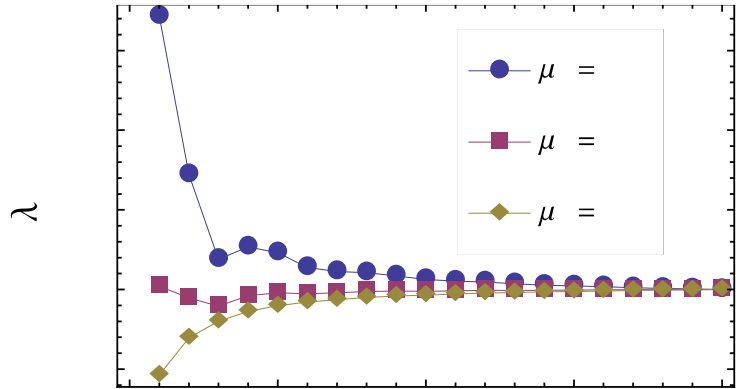

**Supplementary Figure 14. Effective coupling constant $\lambda_{\text{eff}} = \lambda - \mu^*$ as a function of the number of layers $N$, for different values of the bare Coulomb pseudo-potential.** In the graph, three values of the bare pseudo-potential are explored ($\mu_B = 0.1, 0.5, 1.5$) and $\lambda_B = 0.36$. $\lambda_{\text{eff}}$ is normalized to its bulk value for a given set of $\lambda_B$ and $\mu_B$.

**SUPPLEMENTARY NOTES**

**Supplementary Note 1: Optical characterization of 2H-TaS$_2$ flakes.**

The deposition methodology herein described allows for the use of any type of solid substrate as a receiving surface. However, it has already been established how a facile non-destructive detection of atomically thin TMDC layers may be performed by optical microscopy inspection of samples prepared on Si/SiO$_2$ wafers.[1] Preliminary examination of TaS$_2$ patches deposited on a Si wafer with a 285 nm thick SiO$_2$ capping layer permitted the identification of flakes of different thickness flakes as depicted by their different optical contrast (see Supplementary Figure 1).

The presence of this thick silicon dioxide layer between the pure silicon and the deposited material yields an apparent colour that depends on the flake thickness due to a light interference effect.[2] This effect results in very faintly coloured large surface area patches, which correspond to the thinner crystals. In order to quantitatively study the light interference, we measured the thickness dependent optical contrast between the flake and the SiO$_2$ substrate under different illumination wavelengths (see Supplementary Figure 2).[3,4,5,6] For accurate determination of the thickness of the deposited TaS$_2$ flakes, an AFM operated in contact mode was employed. The optical contrast depends both on the flake thickness and the illumination wavelength (shown in Supplementary Figure 2), and can be calculated using a model based on Fresnel laws using the refractive index of TaS$_2$ reported in the literature.[7] We found a significant agreement between the measured optical contrast for thin flakes and that obtained from the model using the refractive index of bulk TaS$_2$. It is remarkable that the optical contrast is strongly dependent on the illumination wavelength and even changes its sign for flakes thinner than 20 nm. This behaviour makes white light illumination inappropriate for the identification of the thinnest flakes by optical microscopy. Oppositely, illumination under certain wavelengths enhances the optical contrast of the thinnest $TaS_2$ crystals, which allowed the optical identification of layers as thin as 1.2 nm.

**Supplementary Note 2: Raman spectroscopy of 2H-TaS$_2$.**

The potential relationship between the flake thickness and the Raman scattering intensity was explored. For this reason a µ-Raman probe was used to explore different thickness flakes. Supplementary Figure 4 shows the thickness dependence of a selection of Raman features. Whereas the frequency shift and the full-width-at-half-maximum (FWHM) of the $A_{1g}$ and $E_{2g}$ Raman modes do not seem to be at all related to the number of layers, it may be clearly appreciated how the ratio between the intensity of the Si peak (at 521 cm$^{-1}$) and the $A_{1g}$ and $E_{2g}$ peaks both increase upon decreasing the number of layers of the probed flake. The frequency difference between the $A_{1g}$ and $E_{2g}$ Raman modes also exhibit a linear proportionality with the number of layers present in the flakes. It is important to highlight that as for other TMDCs, some sensitivity to the Raman laser beam was also exhibited by the TaS$_2$ flakes. In this way, upon performing experiments with long exposure times or high irradiation powers, the flakes were irreversibly damaged as seen by a change in the optical contrast in the focus spot of the laser beam. Yet, no apparent change in the height profile as measured by AFM could be detected. By contrast, the appearance of a strong photoluminescence emission band around 555 nm suggested that some oxidation to Ta$_2$O$_5$ had occurred.[8]

On a final note, it has been previously observed in other transition metal dichalcogenides how the intensity of the distinct Raman modes may vary as the angle between the linearly polarized incident beam and the scattered signal is modified.[9] This can be used to confirm the origin of the Raman peaks. In the TaS$_2$ case, it could be observed that while the intensity of the $E_{2g}$ mode does not depend on the angle between the excitation and detection, the $A_{1g}$ mode presents its maximum intensity for parallel excitation and detection and it vanishes for cross polarized excitation and detection in agreement with that reported for other TMDC flakes,[10] confirming that the Raman signal comes from an analogous crystal (see Supplementary Figure 5).

**Supplementary Note 3. Berezinskii–Kosterlitz–Thouless (BKT) fits to selected devices.**

*I-V* curves were fit to a power law of the form $V \propto I^\alpha$, where α spans from 1 for temperatures above $T_{BKT}$, reaching a value of $\alpha = 3$ at the BKT transition, and monotonically increasing as temperature is further lowered. The R-T curves are further fit to the resistance dependence near the BKT temperature, $R = R_N \exp(-b/(T-T_{BKT})^{1/2})$. Supplementary Figure 7 shows a set of *I-V* and *R-T* curves taken for two devices with thicknesses of 4.7 nm [a – c] and 5.8 nm [d - f]. It may be appreciated how the data do follow a power law with values of the *α* parameter that vary in the expected range typical for 2D superconductivity. From the *α*-exponent analysis and *R-T* transition we estimate a BKT temperature of 1 K for the 4.7 nm flake and 1.2 K for the 5.8 nm flake.

**Supplementary Note 4. Charge density wave (CDW) considerations in 2H-TaS$_2$.**

The experimentally observed charge density wave in 2H-TaS$_2$ has a periodicity of 3 x 3 unit cell in the layer plane. We consider an effective one-orbital tight-binding model and simulate the CDW at mean field level as an onsite potential that locally shift the onsite energy.

The effect of the CDW is seen in the DOS at an energy of 0.2 - 0.3 (in units of the tight binding hopping parameters), in the form of small corrugation arising from the gap opening in part of the band structure (Supplementary Figure 9).

At the Fermi energy, no gap is opened (again visible in Supplementary Figure 9) and the behavior of the DOS at the Fermi level (Supplementary Figure 10), which is ultimately responsible of the $T_c$, is not affected by the presence of the CDW.

**Supplementary Note 5. DFT band structure and tight-binding model.**

A crucial starting point for studying the behavior of the critical temperature versus the thickness of the sample is a faithful description of the system in terms of a band structure and wavefunctions. The details of the calculations are described in the Methods section of the main text.

The resulting band structure of our calculations is shown in Supplementary Figure 11. The bands crossing at the Fermi level have a strong Ta d character. The Fermi surface for a system composed by N layer is constituted by N pockets around $\Gamma$, N pockets in K, and N pockets in K', and in the limit of large N give rise to the well known tubular Fermi surfaces. The calculated total density of states (tot-DOS) is shown in Supplementary Figure 12, where we can see that it presents a large peak slightly below the Fermi level, whose height increase linearly with increasing number of layers, and whose position does not change when varying the number of layers. The tot-DOS at the Fermi level also presents a linear increase with increasing number of layers. From the band structure we understand the peak as arising from van Hove singularities associated to saddle points in the band structure at the M point and at an intermediate point between the K point and $\Gamma$ point. A van Hove peaks is a logarithmic singularity that shows up in the DOS when the system is 2D. As we increase the number of layers the van Hove singularities get smooth and average out when the systems becomes 3D.

The meshing in the DFT simulations, although sufficiently dense for the convergence of the total energy, is not sufficient to resolve the van Hove peak in the DOS. We then construct an effective tight-binding model of a single orbital in a triangular lattice considering in-plane and out-of-plane hopping up to second nearest neighbors,

$$H_0 = -\sum_{n,m=1}^{N} \sum_{i,j} c_{i,n}^* t_j^m c_{i+j,n+m} + H.c.$$

and by fitting the DFT band structure we find $t_1^0 = -0.033$, $t_2^0 = -0.227$, $t_0^1 = -0.039$, $t_1^1 = -0.016$, $t_2^1 = -0.010$, where $t_j^n$ is the hopping matrix element between the n-th out-of-plane and the j-th in-plane nearest neighbor sites. The 2$^{nd}$ in-plane nearest neighbor hopping $t_2^0$ comes out to be larger than the 1$^{st}$ nearest neighbor one, a result that has been already discussed in the literature,[11] and is due to a phase cancelation mechanism, typical of 2H-TMDCs. The resulting DOS per layer at the Fermi level $v_N(0)$ is shown in Supplementary Figure 13, where we clearly see a monotonic decrease of the DOS as we lower the number of layers.

**Supplementary Note 6. Anderson-Morel model.**

The Anderson-Morel model takes into account the effect of a repulsive Coulomb interaction that reduces the effective coupling constant determining the superconducting $T_c$. Here, we generalize the Anderson-Morel model to a system with a generic DOS, that allows us to properly account for the van Hove logarithmic singularities appearing as the systems becomes more and more 2D, as is the case when lowering the thickness of the sample. The idea is to correctly project to the low energy sector the contribution of the high energy repulsive tail of the effective electron-electron interaction. The starting point is a generalized gap equation in the framework of the Eliashberg theory that couples the gap at all energies,

$$\Delta(\varepsilon) = -\int d\varepsilon' V_{\text{eff}}(\varepsilon, \varepsilon') \nu_N(\varepsilon') \frac{\tanh\left(\frac{\varepsilon'}{2T}\right)}{2\varepsilon'} \Delta(\varepsilon')$$

where the effective interaction $V_{\text{eff}}(\varepsilon, \varepsilon')$ is the result of electron-phonon and the electron-electron terms. To keep the problem as simple as possible, we follow the usual treatments and discretize the effective interaction as $\nu_N(0)V_{\text{eff}}(\varepsilon, \varepsilon') = -\lambda + \mu$, for $-\omega_0 \leq \varepsilon, \varepsilon' \leq \omega_0$, and $\nu_N(0)V_{\text{eff}}(\varepsilon, \varepsilon') = \mu$, for $-W \leq \varepsilon, \varepsilon' \leq W$, with W the bandwidth of the DOS. This way, the bandwidth interval is characterized by two regions with different interactions, attractive at a low energy and repulsive at high energy. The gap function can then be separated in two values in the different two regions, $\Delta(\varepsilon) = \Delta$ for $|\varepsilon| < \omega_0$, and $\Delta(\varepsilon) = \Delta\xi$, for $\omega_0 < |\varepsilon| < W$. By introducing the bulk quantities $\lambda_B = V \nu_{\text{bulk}}(0)$, $\mu_B = U \nu_{\text{bulk}}(0)$, the gap equation simplifies to the following system of two coupled equations for $\Delta$ and $\xi$,

$$1 = (\lambda_B - \mu_B)I_1 - \mu_B \xi I_2$$
$$\xi = -\mu_B I_1 - \mu_B \xi I_2$$

where we have defined the following integrals,

$$I_1 = \int_0^{\omega_0} d\varepsilon \frac{\tanh(\varepsilon/2T)}{\varepsilon} \tilde{v}_N(\varepsilon) \qquad I_2 = \int_{\omega_0}^{W} d\varepsilon \frac{\tanh(\varepsilon/2T)}{\varepsilon} \tilde{v}_N(\varepsilon)$$

with $\tilde{v}_N(\varepsilon) = (v_N(\varepsilon) + v_N(-\varepsilon))/2v_{\text{bulk}}(0)$. As we have seen by the DFT simulations and the tight-binding model, the total DOS normalized by the number of layer is featureless close to the Fermi level, so that the integral $I_1$ is performed in the usual way and it gives $I_1 = \ln(1.14\, \omega_0/T)$. At the same time, the DOS displays van Hove singularities at higher energies, that become more and more pronounced as we lower the number of layers. In the limit $T/\omega_0 \ll 1$, the integral $I_2$ can be approximated as

$$I_2 = \int_{\omega_0}^{W} d\varepsilon \frac{\tilde{v}_N(\varepsilon)}{\varepsilon}$$

Using the approximate DOS resulting from the tight-binding model, we checked that the value of $I_2$ monotonically increases with lowering the number of layers, even if the van Hove peaks are smoothed in the numeric approximations. The effective coupling constant $\lambda_{\text{eff}} = \lambda - \mu^*$ is then written as,

$$\lambda_{\text{eff}} = v_N(0) \left( \lambda_B - \frac{\mu_B}{1 + \mu_B\, I_2} \right)$$

where the dependence on the number of layers N is hidden in $v_N(0)$ and $I_2$. As we pointed out in the main text, the renormalization of the pseudo-potential is particularly relevant if the bare Coulomb term is sufficiently strong. In Supplementary Figure 14 we plot the effective coupling constant $\lambda_{\text{eff}}$ for three different values of the bare pseudo-potential, $\mu_B = 0.1, 0.5, 1.5$. Since we are interested in the trend of $\lambda_{\text{eff}}$ with lowering $N$, we choose $\lambda_B$ greater than the bulk value of $1/I_2$, so to guarantee a positive coupling constant. In the computation of $I_2$ we choose the phonon frequency $\omega_0 = 50$ meV and for $\lambda_B = 0.36$. For weak value of $\mu_B$, the effective coupling constant follows the DOS at the Fermi energy, whereas for stronger repulsion we can clearly see that $\lambda_{\text{eff}}$ increases.

## SUPPLEMENTARY REFERENCES